\documentclass[onecolumn,12pt]{IEEEtran}
\usepackage{epsfig}
\usepackage{amsmath}
\usepackage{graphics}
\usepackage{graphicx}

\usepackage{latexsym}
\usepackage{amsfonts}

\renewcommand{\baselinestretch}{1.05}

\begin{document}

\title{On the Achievable Communication Rates of Generalized Soliton Transmission Systems$^*$ \thanks{$^*$Work in progress as part of Eado Meron's PHD} }

\author{Eado Meron, Mark Shtaif  and Meir Feder\\
(e-mail:meroneado@gmail.com)
\thanks{E. Meron M. Shtaif and M. Feder are with the Department of Electrical Engineering-Systems,
Tel Aviv University, Ramat Aviv 69978, Israel
(e-mail:meroneado@gmail.com).}}
\maketitle


\begin{abstract}
We analyze the achievable communication rates of a generalized
soliton-based transmission system for the optical fiber channel.
This method is based on modulation of  parameters of the scattering
domain, via the inverse scattering transform, by the information bits. The decoder uses the direct spectral transform to estimate
these parameters and decode the information message.  Unlike ordinary On-Off Keying (OOK)
soliton systems, the solitons' amplitude may take values in a
continuous interval. A considerable  rate gain is shown in the case
 where the waveforms are 2-bound soliton states.  Using traditional information theory and
inverse scattering perturbation theory, we analyze the influence of
the amplitude fluctuations as well as soliton arrival time jitter,
on the achievable rates. Using this approach we show that the
time of arrival jitter (Gordon-Haus)  limits the information rate in a continuous manner,
 as
opposed to a strict threshold in OOK systems.
\end{abstract}
\section{Introduction}
Communication through optical fiber channels has evolved
enormously in the past couple of decades leading to unprecedented information
rates. Current information
theoretic techniques are unsuccessful in producing  relevant methods to
predict  capacity bounds for these channels.

The nonlinear terms that affect signal evolution led to the following question: \textit{Is the information capacity of the optical fiber channel monotonically increasing with the input power and if so does the capacity grow logarithmically with power as it does for linear channels?}. Moreover, as the complexity allowed in receivers grows, one looks for insights regarding the best (not necessarily the simplest) modulation schemes, signal space and error correcting codes.



The basic generic partial differential  equation (PDE) that describes  the
value of the electric field in space and time (in one dimension)
in the optical fiber channel is (using normalized coordinates and the notations of \cite{Hasegawabook}):
 \begin{equation}\label{nls_loss}
i\frac{\partial q}{\partial Z}+\frac{1}{2} \frac{\partial^2
q}{\partial T^2}+|q|^2q=0
\end{equation}
where the input of the
channel is $q(0,t)$ and the output is $q(L,t)$.  This equation is also known as the non-linear scalar Schrodinger
(NLS) equation.

 Since the equivalent channel is
nonlinear, a Fourier frequency based analysis is not applicable. The
usual way to analyze a continuous-time channel in traditional
information theoretic  methods is to reduce the  problem into a
discrete one by considering the Nyquist samples of the input and
output. However, since a bandlimited input signal evolves into an
output signal of an infinite bandwidth, it is hard to find such
discrete-time models. We stress that the nonlinearity invoked by the
channel is fundamental and is conceptually different than
nonlinearities caused by transmitter/reciever  elements, e.g.,
amplifier nonlinearities, that have been studied in the past.

A different approach to analyzing signal evolution in nonlinear channels
is the inverse scattering transform (IST). In this paper we present this method and apply it to a few tractable problems in which we approximate the achievable data rates. We also explain how this method should be developed to characterize the channel capacity and useful modulation schemes. A similar approach, first proposed by Hasegawa and Nyu (\cite{Hasegawa93}), suggested using multiple solitonic waveforms. It should be noted that the IST approach presented in this paper is not complete in the following aspects:
\begin{itemize}
  \item It does not provide single letter results for capacity but rather a new method to evaluate it which we feel is more esthetic and better suited for this channel.
  \item It does not solve the problems associated with the bounded symbol rate for solitonic waveforms which is characterized by the Gordon-Haus bound (\cite{Gordon86}).
  \item It lacks a simple representation of the manner in which white noise is projected onto complex solitonic waveforms.
\end{itemize}

We now give a short introduction to the inverse scattering transform which solves a set of nonlinear evolution problems via the solution of three linear problems. A recent more complete introduction to the IST and its properties can be found in \cite{nlft1}.
\section{A primer on the inverse scattering transform}
The inverse scattering method does not consist of a single generic transform. In fact, it is more like a recipe for solving a family of nonlinear evolution problems. This recipe involves finding two $q$-dependent operators, $L$ and $M$, that obey certain conditions. The first operator of the two defines an eigenvalue problem for an auxiliary wave function. This problem gives rise to solutions that obey boundary conditions at $-\infty$ and $\infty$. The way these solutions evolve from $-\infty$ to $\infty$ defines the scattering coefficients or the scattering data which is analogous to spectral content in the Fourier frequency domain for linear channel problems. Extracting the scattering data from the $q$ dependent operator is called the direct transform.  Due to special properties of the above operators the evolution of the scattering data in time is rather simple. Moreover, there is a well defined inverse transform that maps the scattering data back to $q$. All of the above steps, direct transform, inverse transform and time evolution are essentially linear problems. We now present the details of the IST for NLS.

To solve integrable systems such as the NLS one needs to  express the system as a compatibility condition of two linear equations for a wave equation, $\Psi(T,Z;\zeta)$:
\begin{eqnarray}\label{corscondition}
L(Z)\Psi & = &\zeta \Psi \label{direct}\\
\frac{\partial \Psi}{\partial(Z)} & = &M(Z)\Psi \label{timeevolution}
\end{eqnarray}
where $L$ and $M$ are differential operators in the $T$-derivatives and are called a Lax pair if:
\begin{equation}\label{laxpair}
\frac{\partial L}{\partial(Z)}=ML-LM\equiv[M,L].
\end{equation}
The right hand side is called the commutator of $M$ and $L$. If (\ref{laxpair})  holds then one can show that the eigenvalues of the operator $L$ are Z-invariant:
$$
d \zeta / dZ=0,
$$
even though $L$ is not $Z$-invariant.

Finding a Lax pair for a given channel is not an obvious task. The Lax pair for the NLS, found by Zacharov and Shabat, is given by:
\begin{eqnarray}\label{zspair}
L  & = & \left(
      \begin{array}{cc}
      i \frac{\partial}{\partial T} & q \\
        -q^* &  -i \frac{\partial}{\partial T} \\
      \end{array}
    \right)\\
     M & = & \left(%
\begin{array}{cc}
  i \frac{\partial^2}{\partial T^2}+\frac{i}{2}|q|^2 & q \frac{\partial}{\partial T}+\frac{1}{2}q_T \\
  -q^* \frac{\partial}{\partial T}+\frac{1}{2}q^*_T  & -i \frac{\partial^2}{\partial T^2}-\frac{i}{2}|q|^2 \\
\end{array}
\right)
     \end{eqnarray}

It is readily verified that for these operators, equation (\ref{corscondition}) results in the NLS equation.
To solve equation \ref{direct} we define vector wave functions for real $\xi=\zeta$ with asymptotic boundary conditions:
\begin{eqnarray}\label{solpair}
                \Phi(T;\xi) & \rightarrow & \left(
                         \begin{array}{c}
                           1 \\
                           0 \\
                         \end{array}
                       \right)e^{-i\xi T} \quad T \rightarrow - \infty \\
                       \Psi(T;\xi)& \rightarrow & \left(
                         \begin{array}{c}
                           0 \\
                           1 \\
                         \end{array}
                       \right)e^{i\xi T} \quad T \rightarrow \infty.
\end{eqnarray}
The pair $\Psi,  \tilde{\Psi}\equiv \{\psi^*_2,-\psi^*_1\}$ is a complete system of solutions for (\ref{direct}). Therefore:
\begin{eqnarray}
\Phi(T,\xi)=a(\xi) \tilde{\Psi}+b(\xi) {\Psi}.
\end{eqnarray}
For $ T \rightarrow \infty$ we have:
\begin{equation}
\Phi(T,\xi) \rightarrow a(\xi)   \left(
                         \begin{array}{c}
                           1 \\
                           0 \\
                         \end{array}
                       \right) e^{-i\xi T} +  b(\xi)   \left(
                         \begin{array}{c}
                           0 \\
                           1 \\
                         \end{array}
                       \right)e^{i\xi T} .
\end{equation}
Comparing with equation (\ref{solpair}) we recognize $1/a(\xi)$ and  $b(\xi)/a(\xi)$ as the transmission and reflection coefficients which characterize the scattering data. The origin of these names is in the fact that they describe what happens to a wave as it evolves from $-\infty$ to $\infty$ and scatters due to a certain "potential", $q$ (these terms are borrowed from quantum physics).

The discrete eigenvalues of the direct scattering problem are the set of points:
\begin{eqnarray}
\zeta=\{\zeta_n,n=1,2,...N;\text{Im}(\zeta)>0 \quad \text{s.t.} \quad a(\zeta)=0\}
\end{eqnarray}
for which:
\begin{eqnarray}\label{directeig}
\Phi(T;\zeta_n)=b_n \Psi(T;\zeta_n).
\end{eqnarray}
Equation (\ref{directeig}) shows that both $\Psi$ and $\Phi$ approach zero as $T$ approaches infinity.
The scattering data, which has a one-to-one correspondence with $q$ and hence carries the same information is comprised of:
\begin{equation}\label{}
    \Sigma(z=0)=[r(\xi;0)=\frac{b(\xi;0)}{a(\xi;0)} \mbox{ for real $\xi$},
    \{\zeta_n,C_n(0)\}\mbox{ for $n=1,2,,,N$}],
\end{equation}
where:
 \begin{equation}
 C_n(0)=b_n(0)/a_n^{'}(0) \quad a_n^{'}(0)= \frac{\partial a}{\partial \zeta}(T=0;\zeta_n)
 \end{equation}
are called the \textit{norming constants} of the bound states.

The time evolution of the scattering data is governed by (\ref{timeevolution}). The solution of which (see \cite{Hasegawabook}) is:
\begin{eqnarray}
r(\xi;Z)& = &r(\xi;0)e^{-i 2 \xi^2 Z}\\
C_n(\zeta_n;Z)& = & C_n(\zeta_n;0)e^{-i 2 \zeta_n^2 Z}\\
 \zeta_n(Z) & = & \zeta_n(0).
\end{eqnarray}

The inverse problem of finding $q$ given the scattering data is solved by a set of linear integral equations which are beyond the scope of this introduction.

 The IST is
important because it allows the use of linear techniques to solve
initial value problems for nonlinear problems.  The main advantages of the IST is
that the number of degrees of freedom that a signal is comprised of, i.e. number of solitons and radiation bandwidth,
does not change through signal evolution and that there are natural invariant-over-time scalar entities, i.e. eigenvalues.
The evolution of the solution in time is most naturally
described through the IST and thus the IST may lead us to insights
regarding communication strategies. For an in-depth survey of the IST also known as the nonlinear Fourier transform, and an OFDM-like communication transmission method, see the paper by Yousefi et al. (\cite{nlft1,nlft2}).

 Actually, Hasegawa and Nyu (see \cite{Hasegawa93,Hasegawabook})proposed
a communication method that utilizes the fact that the eigenvalues
associated with the IST do not change in time. The advantages of the
method proposed by Hasegawa et al. is that it is inherently
multi-valued and is similar to frequency based methods for linear
channels. The authors do not analyze the effects of amplifier noise
on the eigenvalues and its implications on  channel capacity. In the following we
elaborate on the ideas of eigenvalue communications, extend it, and use results
from perturbation theory (see for example \cite{Kaup1976,Kivshar89})
for nonlinear models to estimate the capacity of nonlinear channels. We extend the idea of eigenvalue communication
to that of spectral data modulation and use the inverse scattering transform as our transmitter and the direct spectral transform in the receiver. We quantify the effects of amplitude fluctuations and jitter on
achievable communication rates and evaluate them for realistic
configurations.

\section{Carrying information using the scattering data}
We assume that the channel model is represented by:

\begin{equation}\label{noised_nls}
i\frac{\partial q}{\partial Z}+\frac{1}{2} \frac{\partial^2
q}{\partial T^2}+|q|^2q=\epsilon R
\end{equation}
where $\epsilon R$ is the perturbation term. Throughout this paper
we assume that $R$  is a white noise Gaussian process (in space and time)
with a unit power spectral density (PSD) and  $\epsilon$ is used as
a scaling parameter for the noise power that can be related to the
physical parameters of the channel. We will later plug-in these
parameters to obtain practical results. The noise is generated by
the effects of amplifiers that are spread throughout the fiber but
we assume it is injected adiabatically \footnote{i.e. infinitesimal
noise admitted at every point along the channel} .

The
information rate, $R_b$, that can be achieved on this channel
 is upper bounded  by the channel capacity which is the maximal mutual information between the channel's input and output :\cite{Shannon1948}
\begin{equation}\label{ratebound}
    R_b \leq \max   I(q(0,T);q(L,T)).
\end{equation}
where the maximization is taken over some input constraint (e.g. an
average power constraint, a peak power constraint, Fourier bandwidth or maximal number of solitons).
 Evaluating the quantity above turns out to be a
very difficult task for nonlinear channels. In this paper we argue
that the most tractable  way of evaluating this quantity is through the
statistics of the scattering data of the IST, namely the
eigenvalues and the absolute value of the norming constants.

Since the IST is a one-to-one transformation the mutual
information between the waveforms is equivalent to the mutual
information between the scattering data, i.e.,
\begin{equation}\label{ratebound}
   I(q(0,T);q(L,T))=I(\Sigma(Z=0);\Sigma(Z=L)).
\end{equation}
 To lower bound this quantity one can assume that the input is a
 reflectionless potential so that the information transmitted
 solely through the discrete eigenvalues and corresponding norming
 constants, i.e.,
\begin{eqnarray*}\label{ratebound}
   &&I(\Sigma(Z=0);\Sigma(Z=L)) \geq I(\{\zeta_n(0),C_n(0)\};\{\zeta_n(L),C_n(L)\}) \\
   &&\mbox{ for
   $n=1,2,,,\infty$},
\end{eqnarray*}
where the time index is added since the Gaussian noise changes the
eigenvalues (that are otherwise constant) and can also possibly
change their number via the birth/death of a soliton.

 The observation that the mutual information in
a nonlinear integrable channel can and should be evaluated through the
statistics of the scattering data is the main observation in this
paper. This approach is motivated by several reasons. First,  unlike
the linear spectral domain (i.e., Fourier methods where spectral
broadening is a result of the nonlinearity) the number of degrees of
freedom in the scattering domain remains unchanged throughout the
noiseless evolution. Second, the eigenvalues and norming constants
serve as scalar candidates for the transmission of information
implying a new notion of a nonlinear signal space. The evaluation of equation
(\ref{ratebound}) is  still a cumbersome task, yet it can be
approximated assuming some further restrictions on the input
signals.


\section{Main Results}
In  the generalized soliton transmission system we analyze, a
codeword is a (large) set of symbols. Each symbol is in fact a set
of eigenvalues and norming constants. At
the transmitter, the waveform to be transmitted is generated using
the inverse scattering transform. At the receiver, direct scattering
is applied to derive the set of (perturbed) eigenvalues and norming
constants. The waveforms used by the transmitter have infinite support but decay exponentially so that if we  truncate the waveforms to create a finite symbol period at a suitable distance we can treat the resulting soliton interaction as being negligible to the added noise.


  Throughout this Section the imaginary parts of the
eigenvalues, which can be considered to be generalized amplitudes, will be the information carrying agents.


\begin{figure}[h]
  \centering
  \includegraphics[trim = 0mm 50mm 0mm 70mm, clip,scale=0.4]{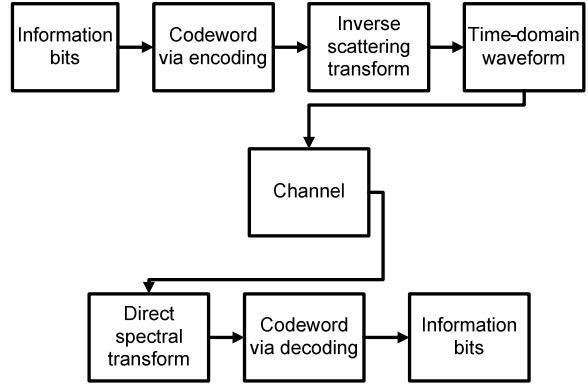}\\
  \caption{The IST-based communication scheme}
  \label{solitoncom}
\end{figure}

\subsection{Information embedded in a single soliton}
In this setting  single
solitons are modulated. Unlike ordinary OOK their amplitudes
belong to a continuous interval.
Without a perturbation, the single soliton solution for the NLS is
\begin{equation}\label{single_soliton}
q(T,Z)= \eta \text{sech} [\eta(T+\kappa Z -T_0)] \exp \left (-i \kappa T +i \frac{\eta^2-\kappa^2}{2}Z+i \sigma_0 \right),
\end{equation}
for which the  corresponding discrete eigenvalue of the IST is
$\zeta=(\kappa+i \eta)/2$. For the rest of the paper we assume all eigenvalues are purely imaginary (except for perturbations).
 The localization of the soliton is around $|T_0|=e^{b(0)\eta}$.



 We use results  from \cite{Hasegawabook} for the first order
perturbations of the eigenvalues. The resulting fluctuation in the amplitude
 is:
\begin{equation}
\frac{d \eta }{dz}= \epsilon \int_\infty^\infty
\Re(R \exp^{-i \varphi}) \text{sech} \tau d \tau
\end{equation}
where $\tau=\eta (T-T_0)$.

Assuming $R$ is a bandlimited white Gaussian noise, i.e.
$<R(t,z),R(\tau,w)>= \delta(t-\tau)\delta(z-w)$, we get:
\begin{equation}
E[(\eta(0)-\eta(Z))^2]=\epsilon^2 \eta  Z
\end{equation}
i.e., the variance of the additive noise is proportional to  $\eta$
(unlike ordinary multiplicative noise for which the variance is
proportional to $\eta^2$).

Thus, assuming information is transmitted in the amplitude of a
single is soliton ($T=0$) we have the following
scalar channel:
\begin{equation}
Y \equiv 2 \Im(\zeta(Z))=\eta(Z)=\eta+\sqrt{\eta}N
\end{equation}
where $N$ is a Gaussian r.v. with zero mean and a variance of $\epsilon^2 Z$. We dismiss the probability that the
soliton vanishes completely and allow for $Y$ to be theoretically zero (or
negative). This scenario can be prevented (with high probability) by
using $\sqrt{\eta}>>\epsilon \sqrt{Z}$ which in the limit of $\epsilon$
going to zero has negligible effect on the capacity. We lower bound
the mutual information for the case $\eta \in [\eta_{min},\eta_{max}]$ with
$\Delta \eta=\eta_{max}-\eta_{min}$. It is  assumed that the noise is
Gaussian and of the the largest possible variance:
 \begin{eqnarray}\label{single_soliton_rate}
 I &\geq& h(Y)-h(Y|\eta)\\
&\geq & h(\eta)-h(Y|\eta)\\ \label{Gaussian_noise}
 &\geq & h(\eta)- \frac{1}{2} \ln 2 \pi e \eta_{max}
\epsilon^2 Z\\
&= &\log \frac{\Delta \eta}{\sqrt{\pi \, e \, \eta_{max} \, \epsilon^2 Z} } \quad
\text{bits/soliton}
\end{eqnarray}
where we use the uniform distribution as the input prior and bound (\ref{Gaussian_noise}) using the fact that Gaussian noise has the highest entropy for a given variance.
We refer to this quantity as the "soliton spectral efficiency" which can be considered to be the NLS analog of spectral efficiency in conventional (linear) channels where it's measured in bits/Hertz.

\begin{figure}[htbp]
  \centering
  \includegraphics[clip,scale=0.5]{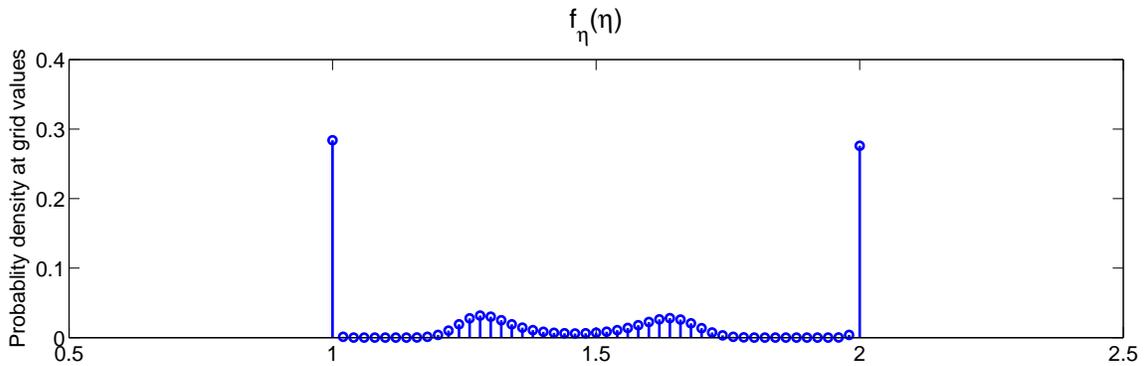}\\
    \caption{$\eta$'s distribution for the square root multiplicative channel}
    \label{capmultsqr}
\end{figure}

\begin{figure}[htbp]
  \centering
 \includegraphics[clip,scale=0.5]{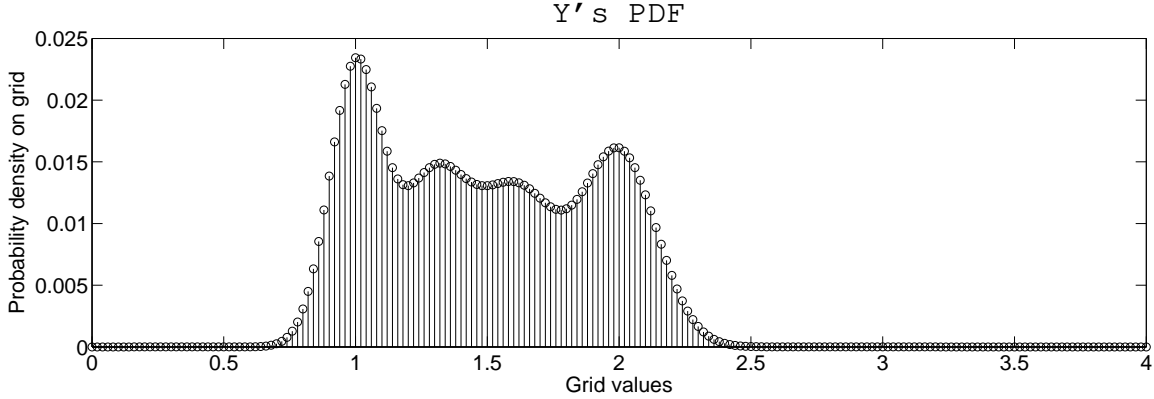}\\
    \caption{Y's distribution for the square root multiplicative channel}
    \label{capmultsqry}
\end{figure}

The capacity can also be directly evaluated using the Blahut-Arimoto algorithm (\cite{Arimoto72,Blahut72}). Using this algorithm for the channel model $Y =\eta+\sqrt{\eta}N$ restricted s.t. $\eta \in [1,2]$ and $E(N^2)=0.1^2$ we get that the true capacity is 1.568 bits per channel use while our bounds reads 1.275 bits per channel use. The capacity achieving prior and the resultant $Y$ distribution are plotted in Figures \ref{capmultsqr} and \ref{capmultsqry}. Note that the capacity achieving prior has both atoms and a continuous distribution which is  typical of interval constrained capacity problems (\cite{Shamai95}).
\subsection{Information embedded in a  soliton train- below the Gordon-Haus rate}

The above result shows that the interval $\Delta \eta=\eta_{max}-\eta_{min}$ should be as large as possible to allow for each soliton to convey as many bits as possible. In fact when one considers transmitting many solitons one after the other, there are other considerations which  bound the optimal interval size from both sides, namely intersoliton interaction and arrival time jitter.

We now consider the case where many solitons are  modulated sequentially. The distance between neighboring solitons is a multiple of the width of the widest soliton, i.e., $\frac{C}{\eta_{min}}$ where $C$ is chosen so that the intersoliton interaction has a negligible (compared to that of the noise) effect on the eigenvalues. The distance between solitons is inversely proportional to the symbol rate and thus in an optimal system $\eta_{min}$ is bounded from below.

 Since we wish to assume a perfectly (or at least an almost perfectly) synchronized communication system, the typical arrival time jitter needs to be less than the distance between neighboring solitons. The time of arrival jitter is known to be directly connected to fluctuations of the real part of the soliton which is linearly related to the velocity of the soliton as can be seen from \ref{single_soliton}. The fluctuations of the real part of the eigenvalue are very similar to that of the imaginary part:
 \begin{equation}\label{kappanoise}
 E[(\kappa(0)-\kappa(Z))^2]= \frac{\epsilon^2 \eta  Z}{3}
 \end{equation}
 Using $\frac{dT_0}{dZ}=-\kappa(Z)$ we integrate to account for the arrival time jitter (neglecting terms that do not originate from the velocity change):
 \begin{equation}\label{arrivaltimenoise}
 E[(T_0(0)-T_0(Z))^2]=\frac{\epsilon^2 \eta  Z^3}{9}
 \end{equation}
 This is the known Gordon-Haus (\cite{Gordon86}) phenomena that bounds the symbol rate of all regular soliton systems (including OOK). The worst-case arrival time jitter is proportional to $\eta_{max}$. Thus, requiring a (almost) jitter free model, e.g., a out-of-synchronization probability of $10^{-9}$ bounds from above $\eta_{max}$.

\begin{figure}[h]
  \centering
   \includegraphics[clip,scale=0.5]{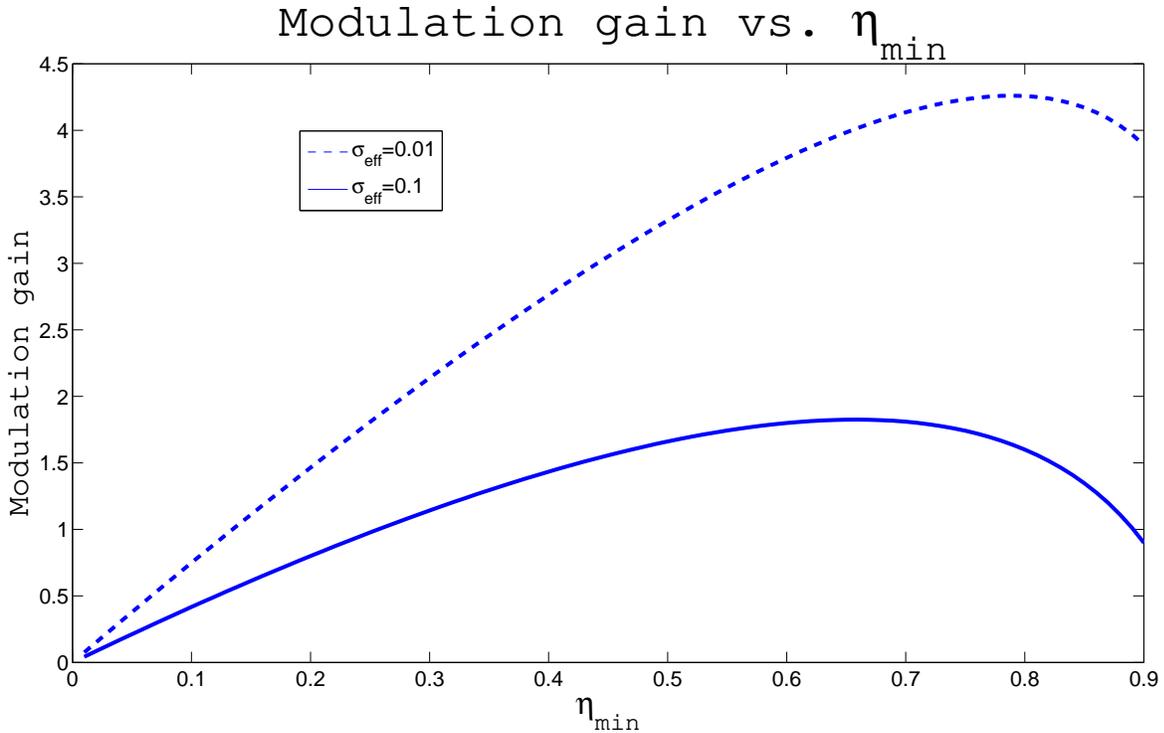}\\
    \caption{Modulation gains as a function of $\eta_{min}$ for different $\sigma_{eff}$.}
    \label{modulationgains1}
\end{figure}

 We wish to compare the gain (in terms of bits/second) of the continuous amplitude modulation scheme versus that of the OOK modulation. We assume that $\eta_{max}$ is tuned by the Gordon-Haus bound requiring no-jitter and is shared by both the continuous system and  the on-off reference system. The continuous system has a lower symbol rate which is $\frac{\eta_{min}}{\eta_{max}}$ times smaller than that of the reference system\footnote{Actually, one can also analyze the case where symbol widths are not constant and are proportional to $1/\eta$}. However, the continuous system conveys more bits than just one per soliton. Weighing both terms the continuous system has a bit rate which is
 \begin{equation} \label{contgain}
 MG=\max_{\eta_{min}} \frac{\eta_{min}}{\eta_{max}} \cdot \log \frac{\Delta \eta}{\sqrt{\pi \, e \, \eta_{max} \, \epsilon^2 Z} } =
 \max_{\eta_{min}} \frac{\eta_{min}}{\eta_{max}} \cdot \log \frac{\Delta \eta}{\sigma_{eff}}
  \end{equation}
 times that of the reference system. We refer to this term as the "Modulation gain". If one would also consider the possibility that a symbol can also contain no soliton at all, and if   $\eta_{min}>>\sqrt{\eta_{max}}\epsilon \sqrt{Z}$ so that the transfer probability between the continuous interval and the zero hypothesis would be less than $10^-3$ than the modulation gain would approximately read:
 \begin{eqnarray} \label{contgainmod}
&& \max_{p} \max_{\eta_{min}} \frac{\eta_{min}}{\eta_{max}} \cdot \left (H_b(p)+ p \cdot \log \frac{\Delta \eta}{\sqrt{\pi \, e \, \eta_{max} \, \epsilon^2 Z} } \right )= \\
 && \max_{p}  \max_{\eta_{min}} \frac{\eta_{min}}{\eta_{max}} \cdot \left (H_b(p)+ p \cdot \log \frac{\Delta \eta}{\sigma_{eff}}   \right ) ,
  \end{eqnarray}
  where $H_b(p)$ is the binary entropy of $p$ (see union of channels in \cite{CoverBook}).
  The modulation gain is plotted in Figure \ref{modulationgains1} for different values of $\sigma_{eff}$. It is evident that as the effective SNR improves a larger $\eta_{min}$ is better since it does not reduce the symbol rate.

\subsection{Information embedded in a  2-bound soliton train- below the Gordon-Haus rate}

The system described above could be analyzed using the framework of perturbations to $\text{sech}$ profiles without necessarily using the perturbation theory of the inverse scattering transform. However, considering more complicated symbols made up of more than one soliton the IST has major analytical and practical advantages. This is the case when the symbols are confined to be either a 2-soliton bound state or a single soliton (or non). We now analyze the modulation gain of this more complicated system and address such issues as common jitter and whether the solitons should be concentric or partially spaced apart.

The idea of transmitting a few concentric solitons is proposed in
the paper by Hasegawa et al. However,  a 2-bound soliton is effected
by noise  differently than each one of its components. We  show
that a 2-bound soliton solution has a larger jitter than its
components. Therefore there is a tradeoff between the enlarged bit
rate and a smaller symbol rate that is induced by a larger jitter.

\begin{figure}[h]
  \centering
   \includegraphics[clip,scale=0.5]{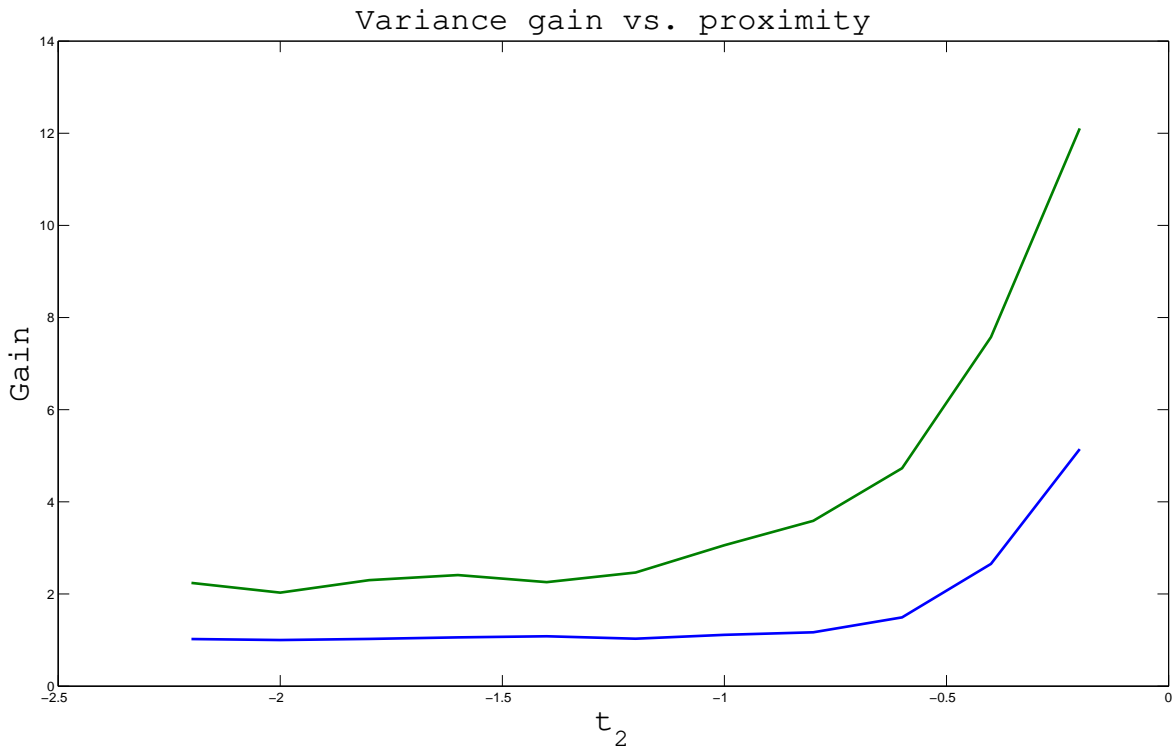}
    \caption{Variance gain (through Monte-Carlo simulations) due to proximity of eigenfunctions. The eigenvalues are 2 and 1. The first eigenfunction has $t=0$ while the second's position is changed.}
    \label{nojitterfig}
\end{figure}

The basic symbol is now comprised of a 2-bound soliton.  This means the transmitter solves the following reflectionless algebraic inverse scattering problem for $N=2$ (\cite{Hasegawabook}):

\begin{eqnarray*}
    f_{ln}&=&\sqrt{C_n}\psi(T;\zeta_n) \quad l=1,2\\
        F_l&=&(f_{l1},,,,f_{lN})\\
        M_{nm}&=&e_n e_m^*/(\zeta_n-\zeta_m^*)\\
        e_n &=& \sqrt{C_n}\exp(i\zeta_n T) \quad E=(e_1,,,,e_N)^t
\end{eqnarray*}
The norming
constants are used to localize the different eigenfunctions. As a
generalization of the single soliton case, we choose $|b_n(0)|=e^{2
\zeta_n t_n(0)}$ where $t_n(0)$ is the generalized position of the
$n_th$ eigenfunction. Actually, the eigenfunctions interact with one
another and the resulting time waveform is not a superposition of 2
single soliton profiles. Nevertheless, their generalized position
remains unchanged throughout the evolution (apart from noise
influence) and can be recovered at the receiver.
The generalized position evolution is given by (to the first order):
\begin{eqnarray*}\label{position_evolution}
t_n(Z)= \left (\frac{\ln b_n(Z)}{\eta_n(Z)} \right)\\
\frac{dt_n(Z)}{dZ}=\kappa(Z)
\end{eqnarray*}
and thus it behaves in the same way as the center of single soliton. However, the fluctuations of the eigenvalues of a 2-bound soliton, both imaginary and real parts are not orthogonal anymore. In fact they are highly correlated in the case of a small separation between generalized locations or in the case of very similar eigenvalues. Moreover, the variance of the fluctuations is generally magnified when the solitons "overlap". This effect makes modulating non-concentric solitons (or actually eigenfunctions) a sensible thing to do. We plot the variance of the eigenvalues as a function of the separation between the generalized positions in Figure \ref{nojitterfig}.
In this setting the detector sees two eigenvalues and two norming constants that translate to generalized positions. All of these scalar quantities are now perturbed by noise. Since the two eigenfunctions are assumed to be much closer to each other than to allow for neglecting the Gordon-Haus jitter, we must account for the way the jitter effects the capacity.

In linear communication problems a non-negligible jitter in symbol
arrival times can diminish the achievable rate to zero. This is
due to the fact that in a linear channel the signal space is made
up of translations of a limited number of base functions. Once
there is a jitter, these functions are no longer orthogonal and
one can not differentiate between neighboring symbols.

However, in a nonlinear  integrable system, solitons can be detected
through the direct scattering transform even if they are one on top
of the other. Actually, they can be detected but not differentiated,
i.e., both will be apparent but the receiver will not know which of
the two belongs to the original slot.

To lower bound the achievable rate of the jitter effected system we
assume that once the eigenfunctions are detected they are sorted
according to time of arrival. This channel is equivalent to
transmitting a couple of solitons (eigenvalues), adding noise and finally
permuting them in the case the switched places. We note the perturbed eigenvalues before  and after the possible permutation $Y_1^n$ and $W_1^n$ correspondingly (n=2 for the 2-bound soliton case). The permutation, which is a random variable, is noted by $\pi_1^n$. The information theoretic loss (in bits) due to the jitter is bounded by:

\begin{eqnarray*}
I(\eta_1^n;Y_1^n)-I(\eta_1^n;W_1^n) \\
& = & h(Y_1^n)-h(W_1^n)-h(Y_1^n|\eta_1^n)+h(W_1^n|\eta_1^n) \\
& \leq& h(W_1^n|\eta_1^n)-h(Y_1^n|\eta_1^n) \\
 & \leq & h(Y_1^n,\pi_1^n|\eta_1^n)-h(Y_1^n|\eta_1^n)\\
 & = &H(\pi_1^n|Y_1^n,\eta_1^n) \\
& \leq & H(\pi_1^n)
\end{eqnarray*}

For the two soliton case, the permutation R.V. is equivalent to a Bernoulli R.V. where the mix-up probability is equal to the probability that the order of the generalized positions is changed. Using the assumption that the eigenvalues will approximately fluctuate in the same way as if the solitons were apart (and this is not true when they walk-by each other) we can approximate this probability. For the set of generalized positions {-1,1}, this probability is equal to $p_{mix-up}<\Pr (\Delta T>1)$ where $\Delta T \sim N(0,\frac{\epsilon^2 \eta_{max}  Z^3}{9})$. If this probability turns out to be $p_{mix-up}=0.1$, which is conventionally thought to be prohibitively large, the rate loss is only $H_b(0.1) \approx 0.5$ bits for the 2-soliton symbol and only 0.25 bits per soliton ($H_b(p)$ is the Shannon binary entropy function). The main advantage is a major increase in the soliton rate, since there are two solitons per symbol.

\begin{figure}[h]
  \centering
   \includegraphics[clip,scale=0.5]{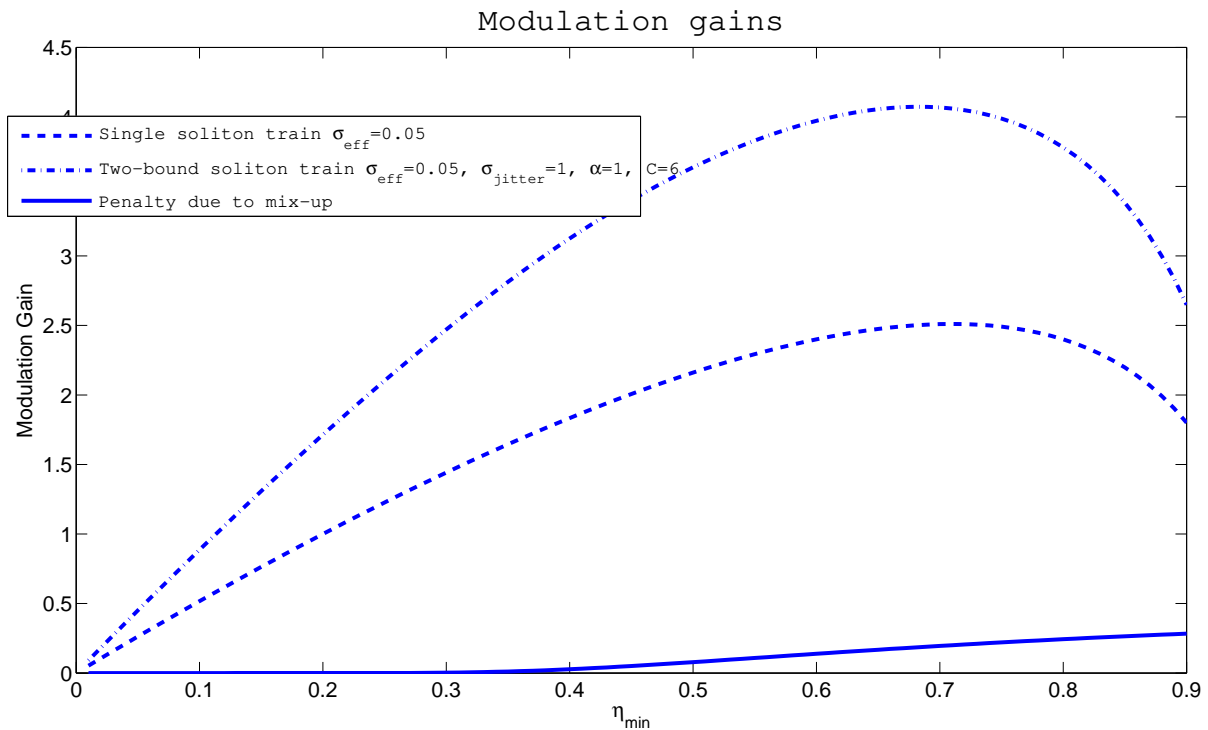}
    \caption{Modulation gains as a function of $\eta_{min}$ for two soliton trains vs. single soliton trains.}
    \label{modulationgains2}
\end{figure}

Assuming the spacing between solitons of the same symbol is about $\alpha/\eta_{min}$ and that original distance between symbols was $C/\eta_{min}$ the soliton rate is increased by a factor of $2 \cdot \frac{C}{C+1}$. We approximate the mix-up probability to be $p_{mix-up} \approx Q \left (\frac{\alpha/\eta_{min}}{\sigma_{jitter}} \right )$. Thus for this setting the "modulation gain" compared to a simple OOK system is approximately:
\begin{equation} \label{cont2gain}
\max_{p} 2 \cdot \frac{C}{C+\alpha} \max_{\eta_{min}} \frac{\eta_{min}}{\eta_{max}} \cdot \left (H_b(p)+ p \log \frac{\Delta \eta}{\sqrt{\pi \, e \, \eta_{max} \, \epsilon^2 Z} }-H_b(p_{mix-up})/2 \right ) .
  \end{equation}
The modulation gain for a certain set of parameters is shown in Figure  \ref{modulationgains2} . The gain compared to single soliton trains is roughly 2 for a wide set of parameters.

\subsection{Approximating the Information embedded in a soliton train- slightly above Gordon-Haus rate}
The next natural generalization is to consider an N-bound solution that is made up a train of well-spaced (spacing relates to the value of the norming constants) eigenfunctions (we assume N to be large, i.e. >5). The analysis of the former subsection is still a good approximation. The difference is that now the ambiguity in time of arrival is not bounded to a pair of solitons. Still, if the eigenfunctions are properly spaced the entropy of the order-of-arrival sequence, $H(\pi_1^n)$, is mainly to do with the probability that consecutive eigenfunctions will change their order of arrival. The information theoretic penalty on the bit rate due to this effect is:
\begin{equation}
\frac{1}{2} H(p_{mix-up}, 1-2p_{mix-up}, p_{mix-up}) \approx p_{mix-up} \cdot \log{1/p_{mix-up}}.
\end{equation}
Now, assume the spacing between solitons is approximately $\frac{1}{\eta_{min}}$ (much smaller than the one called for by the Gordon-Haus limit) and the total modulation gain in this setting is:
\begin{equation} \label{contNgain}
  \max_{p}\max_{\eta_{min}} \frac{\eta_{min}}{\eta_{max}} \cdot \left (H_b(p)+ p \log \frac{\Delta \eta}{\sqrt{\pi \, e \, \eta_{max} \, \epsilon^2 Z} }-p_{mix-up} \cdot \log{1/p_{mix-up}} \right ) .
  \end{equation}

Again, there is no problem with trains of eigenfunctions with a typical mix-up (between consecutive eigenfunctions) probability of $0.1$. Moreover now there is a clear tradeoff for eigenfunction spacing. The bigger the spacing the smaller the symbol rate. As the spacing becomes smaller the penalty due to jitter is larger and so a unique maximum exists. The main disadvantage compared to the previous subsection is that the processing now involves a more complicated channel code. The main advantage is a larger symbol rate.

The analysis above neglects a few things:
\begin{enumerate}
    \item There is small coupling between amplitude and
    time-of-arrival fluctuations. A precise analysis should only
    yield a higher rate.
    \item     When two solitons pass by  each other, their perturbation
    statistics is changed. In many cases, their amplitude
    fluctuations grow and are now dependent. We ignore the growth
    in fluctuations since, assuming that solitons are not too
    crowded, the walk-off is time bounded and its effects are
    negligible. Furthermore, the dependency can only increase the
    rate.
    only
    \item We ignore the possibility that a soliton will die/be
    born. This happens with a small probability and we assume
    that its effect on the achievable rates can also be bounded.
\end{enumerate}

\section{Discussion and further work}
The notion of modulating the ``natural" domain of the channel is not new to communication theory. In fact, the scheme discussed in this paper can be considered to be the nonlinear analog of OFDM. Both of the methods allow for a natural examination of their respective channel capacities. There are two main differences between the two methods. The first is that in linear channels the noise projection on different modes (spectral bands) is orthogonal while in the nonlinear case the noise projection on different modes (solitons) is orthogonal only in some cases (see Figure ). The second is that OFDM is very efficient in terms of complexity (through the use of the celebrated FFT and IFFT) while the direct scattering is a computationally intensive method.

Future research directions include:

\begin{enumerate}
\item Find reasonable complexity (preferably analog) methods to carry out the tasks of  inverse and especially direct scattering in the transmitter and receiver.
    \item Use the approach discussed in the paper with more complex potentials/waveforms (not reflection-less) to lower and upper bound the overall capacity (and not just achievable rates).
  \item While the problems above are not related to information theory, there is a totally new and interesting information-theoretic problem that relates to communication via the scattered domain. When receiving waveforms that are comprised of N-bound solitons or solitons that are co-centric due to jitter (and not thru the constructed modulation) one detects a set of scalar values that can be detected but not differentiated. Essentially, the transmitter and receiver communicate through the transmission of a set, not a sequence, of perturbed scalar values. Clearly, transmitting and receiving a 3-bound solitons conveys less information than a sequence of (ordered in time) three solitons. The question is how much less?
      We call this problem: \textbf{communicating with colorless, but not massless, balls}.  For more on this issue see the work by Meron et al. \cite{ITset}.

\end{enumerate}

%

\bibliographystyle{IEEEtran}
\bibliography{IEEEabrv,refs4}

\end{document}